Letter

# Imaging ferroelectric domains with soft-x-ray ptychography at the oxygen K-edge

Tim A. Butcher [1,2,*] Nicholas W. Phillips [1,3] Chia-Chun Wei [4] Shih-Chao Chang [4]
Igor Beinik [5] Karina Thånell [5] Jan-Chi Yang [4,6] Shih-Wen Huang [1] Jörg Raabe [1] and
Simone Finizio [1]

[1] *Paul Scherrer Institut, 5232 Villigen PSI, Switzerland*
[2] *Max Born Institute for Nonlinear Optics and Short Pulse Spectroscopy, 12489 Berlin, Germany*
[3] *Minerals Resources CSIRO, Clayton, Victoria 3168, Australia*
[4] *Department of Physics, National Cheng Kung University, Tainan 70101, Taiwan*
[5] *MAX IV, Lund University, 224 84 Lund, Sweden*
[6] *Center for Quantum Frontiers of Research & Technology (QFort), National Cheng Kung University, Tainan 70101, Taiwan*

The ferroelectric domain structure of a freestanding $BiFeO_3$ film was visualized by ptychographic dichroic imaging with linearly polarized x-rays at the O K-edge around 530 eV. The dichroic contrast is maximized at the energy of the hybridization of the O 2p state and the Fe 3d orbitals, which is split by the octahedral crystal field of the perovskite structure. The microscopy images thus obtained complement the ptychographic imaging of the antiferromagnetic contribution at the Fe $L_3$-edge. The approach can be extended to the separation of different ferroic contributions in other multiferroic oxides.

Transition-metal oxides can host magnetic and ferroelectric orders. Their magnetic ordering is superexchange mediated, which gives rise to either antiferromagnetism or ferrimagnetism. Ferroelectricity originates in the lone-pair mechanism, in charge ordering, or it is driven either geometrically or by spin [1]. Among the most prominent oxides with perovskite structure is bismuth ferrite ($BiFeO_3$; BFO), which is a room-temperature multiferroic. The $Fe^{3+}$ cations are coordinated octahedrally by six $O^{2-}$ anions, which is shown in the crystal structure in Fig. 1(a). The ferroelectric Curie temperature $T_C = 1110$ K almost reaches the temperature of decomposition, while the antiferromagnetic Néel temperature lies at $T_N = 640$ K. The ferroelectric polarization is parallel to the pseudocubic $\langle 111 \rangle$ directions and arises from the lone-pair mechanism. Nonstrained BFO shows noncollinear order with a Dzyaloshinskii-Moriya interaction-mediated spin cycloid that propagates orthogonally to the ferroelectric polarization with a period of approximately 64 nm [2,3].

Microscopy methods are imperative for the visualization of ferroelectric domain structures, and those involving x-rays from synchrotrons have become indispensable for the study of ferroic materials. The strengths of x-ray microscopy are twofold: the sensitivity to ferroic orders by dichroic contrasts and the elemental sensitivity to various absorption edges. Ferromagnetism is accessible by means of x-ray magnetic circular dichroism, which is sensitive to the magnetization component pointing along the wave vector of the incident x-ray beam. Detection of ferroelectricity and antiferromagnetism both rely on x-ray linear dichroism (XLD), which is sensitive to the respective component perpendicular to the wave vector of the probing x-ray beam. As the angular dependence of the XLD is identical for both antiferromagnetic and ferroelectric ordering, the separation of their contributions in multiferroic materials is not straightforward when measuring at the absorption edge of the magnetic transition metal that is sensitive to both. In the case of periodic noncollinear magnetic ordering such as in BFO, it is possible to sidestep the issue by resolving the modulation of the magnetic order alongside the ferroelectric domains with a microscopy method with sufficiently high spatial resolution. This was recently achieved for the spin cycloid and the ferroelectric domains in BFO with soft-x-ray ptychography at the Fe $L_3$-edge (707 eV) [4]. However, imaging at the absorption edge of the nonmagnetic element is unavoidable when the antiferromagnetism of the

---

multiferroic material is collinear. Hitherto, this has only been achieved at the Ti $L_3$-edge (457 eV) [5–7] and the O K-edge (530 eV) [8] with the surface- sensitive technique of x-ray photoemission electron microscopy (XPEEM). However, these measurements were only capable of spatial resolutions in the order of 70 nm. Higher spatial resolutions and bulk sensitivity are obtainable with coherent diffractive imaging (CDI) methods such as holographically assisted CDI [9,10] or the aforementioned soft-x-ray ptychography [11]. Ptychographic imaging relies on scanning the sample through the x-ray beam while recording diffraction patterns along the trajectory [12]. An iterative phase retrieval algorithm is used to reconstruct the diffraction data to obtain a real-space image of the sample. The origin of XLD is the angular-dependent absorption of linearly polarized x-rays, which is caused by anisotropic charge distributions in the sample plane and appears in the reconstructed ptychographic images. Dichroic ptychography at the O K-edge was first used for a polarization dependent study of $CaCO_3$ crystallites [13]. The following is a demonstration that soft-x-ray ptychography can provide high quality linear dichroic images of ferroelectric domains at the O K-edge with the case study of BFO.

A freestanding 80-nm-thin film of BFO was chosen for this study. The BFO film was grown along the (001) direction by pulsed laser deposition on a [001]-oriented $SrTiO_3$ substrate with an intermediate sacrificial layer of $Sr_3Al_2O_6$ (SAO) as detailed in Ref. [14]. The SAO was then dissolved in water to obtain the freestanding film, which was transferred to a transmission electron microscopy (TEM) grid with a lacy carbon support.

An x-ray transmission spectrum was measured at the O K-edge with scanning transmission x-ray microscopy (STXM) and is displayed in Fig. 1(b). The O K-edge covers an energy range of 30 eV and is caused by the excitation of the O 1s core level to the O 2p state. Its structure for the perovskite BFO is representative for transition-metal oxides. The shape of the O K-edge in perovskites is influenced by the hybridization of the O 2p states and the valence states of the metallic constituents [15]. The measured spectrum can be divided into three contributions that are shaded in Fig. 1(b). The first two peaks stem from the hybridization with the Fe 3d states that are split due to the octahedral crystal field into $t_{2g}$ at 530 eV and $e_g$ around 531.5 eV [16,17]. The absorption is stronger at the $e_g$ peak than the $t_{2g}$ peak, which is in accordance with the structure of the spin-split Fe $L_3$-edge [18]. The absorption peak at 533.5 eV is assigned to the hybridization with Bi 6sp states and constitutes the second region. The third region consists of the broad peak centered at 540 eV and is attributed to hybridization with Fe 4sp states.

The ferroelectric polarization in BFO is parallel to its $\langle 111 \rangle$ direction and the XLD effect is sensitive to its in-plane component when the x-rays are under normal incidence with respect to the sample plane. For maximization of the XLD, the BFO film was aligned with the [110]-axis within $\theta = 5°$ of the oscillating electric field of the inbound soft x-rays. Piezoresponse force microscopy (PFM) evinces that the ferroelectric polarization directions change by 71° at the majority of domain walls in 80-nm-thin freestanding films [4]. This means a 90° change of the in-plane projection of the ferroelectric polarization **P** between domains, which maximizes the XLD contrast.

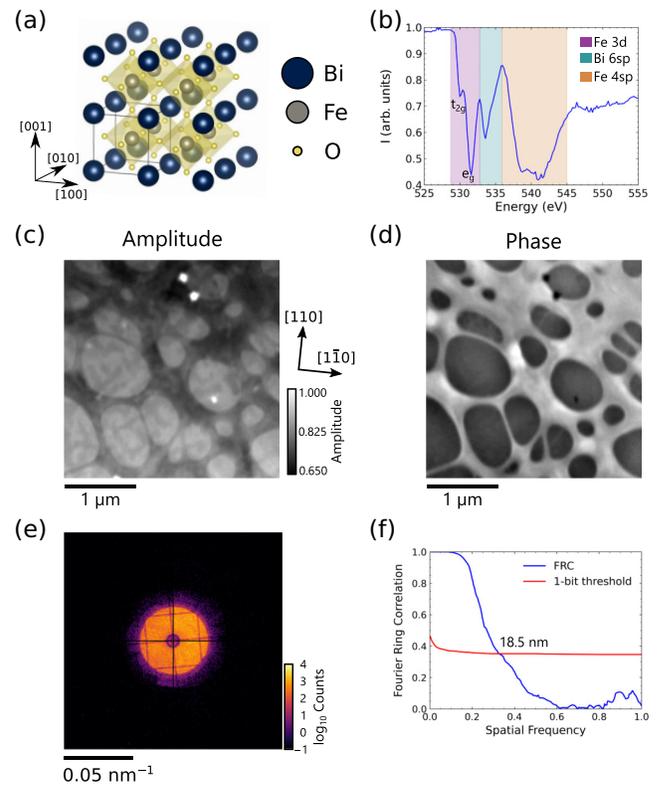

FIG. 1. (a) Perovskite crystal structure of BFO. The $Fe^{3+}$ cations (grey) are octahedrally coordinated by the $O^{2-}$ anions (yellow). (b) O K-edge x-ray transmission spectrum of BFO. The hybridization of the O 2p states with Fe and Bi leads to three sections containing absorption features. The first begins at 529 eV with the Fe 3d states that are crystal field split into $t_{2g}$ and $e_g$. The peak at 533.5 eV is caused by hybridization with Bi 6sp states. Hybridization with Fe 4sp states leads to the broad absorption peak around 540 eV. (c),(d) Ptychographic images at the $e_g$ peak of the O K-edge (531.5 eV) with horizontally polarized x-rays. The ferroelectric domains are visible as darker and lighter regions in the amplitude image overlaying the main absorption due to the lacy carbon support, which dominates the phase image. (e) Exemplary diffraction pattern collected with the single-photon-counting LGAD Eiger detector. (f) A spatial resolution of 18.5 nm was determined from the FRC with the 1-bit criterion.

Ptychography scans were carried out at the energies of the first three peaks in the O K-edge. The detector was a $512 \times 512$ pixel single-photon-counting low-gain avalanche diode (LGAD) Eiger soft-x-ray detector with a square pixel size of 75 μm. The sample-detector distance was 96 mm and the movement of the sample during the ptychographic scan followed a Fermat spiral [19] with a step size of 200 nm. This resulted in 624 scanning points for a 5 μm × 5 μm area. The sample was positioned by means of a piezoelectric stage that was coupled with a differential heterodyne laser interferometer to correct for uncertainties due to vibrations. Diffraction patterns were recorded at every step with an exposure time of 200 ms and a 900-nm FWHM beam on the sample. These were reconstructed with the PTYCHOSHELVES software [20] using 1000 difference-map iterations [21]. The reconstruction algorithm was initiated with the illumination (probe) that was obtained from the ptychographic reconstruction of a strongly scattering sample at a photon energy of 700 eV. Three probe modes were used to account for the presence of partial coherence [22].

The resulting amplitude and phase images for the scan at the $e_g$ peak at 531.5 eV are shown in Figs. 1(c) and 1(d). The lacy carbon of the substrate is immediately visible with its contrast inverting between amplitude and phase images. Inspecting the amplitude image, a mosaiclike pattern of ferroelectric domains in lighter and darker areas can be discerned. Such an irregular domain pattern was previously observed in 80-nm- thin BFO films at the Fe $L_3$-edge [4], which corresponds to the crystal field split $e_g$ states in which the 3d orbitals point towards the O atoms [18]. The domains were also perceivable with ptychography at the peak induced by hybridization with Bi 6sp states, although they were more faint with weaker XLD contrast. The $t_{2g}$ peak of the O K-edge spectrum showed no XLD in the ptychographic reconstructions.

An exemplary diffraction pattern from the scan leading to Figs. 1(c) and 1(d) can be viewed in Fig. 1(e). This shows the first diffraction order from the Fresnel zone plate as measured in the far field, additionally to the scattered photons from the sample. The spatial resolution of the image was estimated by means of a Fourier ring correlation (FRC), which is shown in Fig. 1(f). The FRC was computed for a duplicate ptychography scan with identical scan parameters. This indicated a value of 18.5 nm for the spatial resolution according to the 1-bit criterion, which is comparable to a high-resolution STXM or PFM measurement. The LGAD Eiger suffers from a decreased quantum efficiency at energies below 550 eV, and this can explain the inefficient detection of scattering at high $q$-values, which limits the achievable spatial resolution.

Separation of the pure ferroelectric XLD contrast from the sample topography was achieved by subtracting the amplitude images recorded with horizontally and vertically polarized x-rays, the result of which is shown in Fig. 2 for a different area than in Fig. 1(c). This XLD amplitude image at the O K-edge is proportional to **P** and shows the mosaiclike domain pattern. The geometry of the sample with respect to the linear polarizations of the x-rays is sketched in the inset. It is impossible to directly recover the direction of the ferroelectric polarization vector exclusively from the XLD image, as the effect has a $1 - \cos^2\theta$ dependence [5,13]. It follows that 180° domains are indistinguishable with XLD as they show identical contrasts. A previous PFM study of the ferroelectric domains in freestanding BFO films between paraelectric SrRuO$_3$ (SRO) in a Pt/SRO/BFO/SRO stack reported a predominance of large 180° ferroelectric domains [23]. The absence of 180° domains in our freestanding films may be related to the presence of ionic adsorbates at the surface that compensate the depolarizing field that would otherwise favor 180° domains for the minimization of electrostatic energy [24]. The electrostatic boundary conditions at the BFO/SAO interface prior to the liftoff of the freestanding BFO film are tunable by adjusting the oxygen deposition pressure. Hence, the ion or electron adsorbates at the interface can be controlled and the presence of 71° or 180° domains in the freestanding film adjusted.

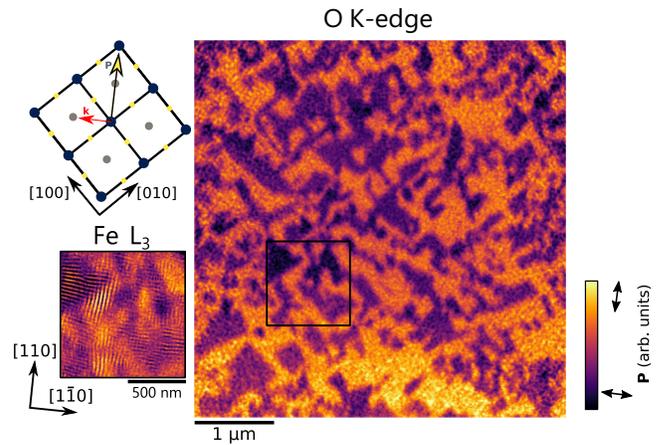

FIG. 2. Linear dichroic contrast image with ferroelectric domains. The amplitude XLD contrast image at the energy corresponding to hybridization of the O 2p with the Fe 3d $e_g$ states (531.5 eV) of the O K-edge shows the pure in-plane projection of the ferroelectric polarization **P** without the sample topography. The bottom inset shows the resulting XLD phase image at the $e_g$ peak of the Fe $L_3$-edge (707.5 eV), which includes the antiferromagnetic spin cycloid. The top inset shows the geometry of the crystal axes, in-plane projection of the ferroelectric polarization **P** and cycloid propagation direction **k** for one domain.

In order to demonstrate the utility of the elemental sensitivity of the x-rays, a ptychographic XLD image was obtained at the x-ray energy of the second of the crystal field split Fe $L_3$ peaks (707.5 eV). This XLD image at the Fe $L_3$-edge is displayed in the inset of Fig. 2 for the area located within the black square. It shows the stripes due to the antiferromagnetic spin cycloid superimposed on the ferroelectric domains that are also present in the image at the O K-edge. The 80 nm thin freestanding film is unstrained and the cycloid appears with a periodicity of 64 nm as in bulk crystals of BFO [3]. The cycloid appears with half its period due to the square dependence of the linear dichrorism [4]. In the case of freestanding BFO with 80 nm thickness, the propagation direction of the cycloid **k** lies within the plane of the film along the ⟨110⟩ directions that are orthogonal to **P** (see sketch of sample geometry in the top inset of Fig. 2). The soft-x-ray ptychography measurements directly prove the strong magnetoelectric coupling that enforces the change of the cycloid propagation direction at the domain walls in unison with the change of ferroelectric polarization.

In conclusion, the sensitivity of soft x-rays to ferroic order coupled with elemental selectivity measurements allows full characterization of multiferroic properties with a single technique. Despite the challenging energy range for coherent diffractive imaging, soft-x-ray ptychography at the O K-edge provides a versatile and efficient way to map ferroelectric domains and directly compare them to magnetic domain patterns in the case of multiferroic compounds such as BFO. Thus, it is possible to directly visualize the magnetoelectric coupling in multiferroic oxides, which avoids the change of microscopy methods and cumbersome search for the originally analyzed area this entails [25]. It was shown that the structure of the O K-edge, which is strongly influenced by the hybridization with metallic cations, is key to the XLD of the compound. The results for BFO can be extended to other perovskites with their octahedral coordination of a metal cation by the $O^{2-}$ anions and the ensuing crystal field splitting that provides sensitivity of the orbitals to asymmetries in the charge distribution due to ferroelectric polarization. This is particularly welcome in the case of multiferroics with collinear antiferromagnetism for which measurements at the absorption edge of the magnetic element have the same contrast as the ferroelectric XLD. Such a situation can arise in BFO, for instance due to strain [26] or substitution of Bi by La [27]. More generally, imaging at the O K-edge is also valuable for the study of other ferroelectrics that contain heavy elements with 4d or 5d electrons and no x-ray absorption edges in the soft-x-ray range. Further improvements in the spatial resolution of soft-x-ray ptychography can be expected with the development of soft-x-ray detectors and complementary characterization of samples at the atomic scale is enabled by TEM.

*Acknowledgments.* Soft-x-ray ptychography measurements were performed with the SOPHIE endstation at the SoftiMAX beamline of the MAX IV Laboratory. The SOPHIE endstation was designed and assembled at the Swiss Light Source (SLS), Paul Scherrer Institut (PSI), Villigen, Switzerland. Research conducted at MAX IV, a Swedish national user facility, is supported by the Swedish Research Council under contract 2018-07152, the Swedish Governmental Agency for Innovation Systems under contract 2018-04969, and Formas under contract 2019-02496.

Preliminary ptychographic imaging was performed at the Surface/Interface Microscopy (SIM-X11MA) beamline of the SLS, PSI, Villigen, Switzerland. The sample was precharacterized at the surface diffraction endstation at the MS (X04SA) beamline of the SLS, PSI, Villigen, Switzerland. T.A.B. acknowledges funding from the Swiss Nanoscience Institute (SNI) and the European Regional Development Fund (ERDF). N.W.P. received funding from the European Union's Horizon 2020 research and innovation program under the Marie Skłodowska-Curie grant agreement no. 884104. J.-C.Y. acknowledges the financial support from National Science and Technology Council (NSTC) in Taiwan, under grant no. NSTC 112-2112-M-006-020-MY3 and no. NSTC 113-2124-M-006-010. We thank E. Fröjdh, F. Baruffaldi, M. Carulla, J. Zhang, and A. Bergamaschi for the development of the LGAD Eiger detector. The LGAD sensors were fabricated at Fondazione Bruno Kessler (Trento, Italy).

[1] M. Fiebig, T. Lottermoser, D. Meier, and M. Trassin, The evolution of multiferroics, Nat. Rev. Mater. **1**, 1 (2016).

[2] I. Sosnowska, T. P. Neumaier, and E. Steichele, Spiral magnetic ordering in bismuth ferrite, J. Phys. C: Solid State Phys. **15**, 4835 (1982).

[3] S. R. Burns, O. Paull, J. Juraszek, V. Nagarajan, and D. Sando, The experimentalist's guide to the cycloid, or noncollinear antiferromagnetism in epitaxial BiFeO₃, Adv. Mater. **32**, 2003711 (2020).

[4] T. A. Butcher, N. W. Phillips, C.-C. Chiu, C.-C. Wei, S.-Z. Ho, Y.-C. Chen, E. Fröjdh, F. Baruffaldi, M. Carulla, J. Zhang, A. Bergamaschi, C. A. F. Vaz, A. Kleibert, S. Finizio, J.-C. Yang, S.-W. Huang, and J. Raabe, Ptychographic nanoscale imaging of the magnetoelectric coupling in freestanding BiFeO₃, Adv. Mater. **36**, 2311157 (2024).

[5] S. Polisetty, J. Zhou, J. Karthik, A. R. Damodaran, D. Chen, A. Scholl, L. W. Martin, and M. Holcomb, X-ray linear dichrorism dependence on ferroelectric polarization, J. Phys.: Condens. Matter **24**, 245902 (2012).

[6] R. V. Chopdekar, V. K. Malik, A. Fraile Rodríguez, L. Le Guyader, Y. Takamura, A. Scholl, D. Stender, C. W. Schneider, C. Bernhard, F. Nolting, and L. J. Heyderman, Spatially resolved strain-imprinted magnetic states in an artificial multiferroic, Phys. Rev. B **86**, 014408 (2012).

[7] M. Ghidini, F. Maccherozzi, X. Moya, L. C. Phillips, W. Yan, J. Soussi, N. Métallier, M. E. Vickers, N. J. Steinke, R. Mansell, C. H. W. Barnes, S. S. Dhesi, and N. D. Mathur, Perpendicular local magnetization under voltage control in Ni films on ferroelectric BaTiO₃ substrates, Adv. Mater. **27**, 1460 (2015).

[8] R. Moubah, M. Elzo, S. El Moussaoui, D. Colson, N. Jaouen, R. Belkhou, and M. Viret, Direct imaging of both ferroelectric and antiferromagnetic domains in multiferroic BiFeO₃ single crystal using X-ray photoemission electron microscopy, Appl. Phys. Lett. **100**, 042406 (2012).

[9] A. S. Johnson *et al.*, Ultrafast X-ray imaging of the light-induced phase transition in VO₂, Nat. Phys. **19**, 215 (2023).

[10] L. Vidas, C. M. Günther, T. A. Miller, B. Pfau, D. Perez-Salinas, E. Martínez, M. Schneider, E. Gührs, P. Gargiani, M. Valvidares, R. E. Marvel, K. A. Hallman, R. F. J. Haglund, S. Eisebitt, and S. Wall, Imaging nanometer phase coexistence at defects during the insulator–metal phase transformation in $VO_2$ thin films by resonant soft X-ray holography, Nano Lett. **18**, 3449 (2018).

[11] D. A. Shapiro, Y.-S. Yu, T. Tyliszczak, J. Cabana, R. Celestre, W. Chao, K. Kaznatcheev, A. D. Kilcoyne, F. Maia, S. Marchesini *et al.*, Chemical composition mapping with nanometre resolution by soft X-ray microscopy, Nat. Photonics **8**, 765 (2014).

[12] F. Pfeiffer, X-ray ptychography, Nat. Photonics **12**, 9 (2018).

[13] Y. H. Lo, J. Zhou, A. Rana, D. Morrill, C. Gentry, B. Enders, Y.-S. Yu, C.-Y. Sun, D. A. Shapiro, R. W. Falcone, H. C. Kapteyn, M. M. Murnane, P. U. P. A. Gilbert, and J. Miao, X-ray linear dichroic ptychography, Proc. Natl. Acad. Sci. U.S.A. **118**, e2019068118 (2021).

[14] D. Lu, D. J. Baek, S. S. Hong, L. F. Kourkoutis, Y. Hikita, and H. Y. Hwang, Synthesis of freestanding single-crystal perovskite films and heterostructures by etching of sacrificial water-soluble layers, Nat. Mater. **15**, 1255 (2016).

[15] F. Frati, M. O. J. Y. Hunault, and F. M. F. de Groot, Oxygen K-edge X-ray absorption spectra, Chem. Rev. **120**, 4056 (2020).

[16] K.-T. Ko, M. H. Jung, Q. He, J. H. Lee, C. S. Woo, K. Chu, J. Seidel, B.-G. Jeon, Y. S. Oh, K. H. Kim, W.-I. Liang, H.-J. Chen, Y.-H. Chu, Y. H. Jeong, R. Ramesh, J.-H. Park, and C.-H. Yang, Concurrent transition of ferroelectric and magnetic ordering near room temperature, Nat. Commun. **2**, 567 (2011).

[17] C.-S. Chen, C.-S. Tu, P.-Y. Chen, W. S. Chang, Y. U. Idzerda, Y. Ting, J.-M. Lee, and C. W. Yu, Micro-to-nano domain structure and orbital hybridization in rare-earth-doped $BiFeO_3$ across morphotropic phase boundary, J. Am. Ceram. Soc. **101**, 883 (2018).

[18] Y. Ting, C.-S. Tu, P.-Y. Chen, C.-S. Chen, J. Anthoniappen, V. H. Schmidt, J.-M. Lee, T.-S. Chan, W.-Y. Chen, and R.-W. Song, Magnetization, phonon, and X-ray edge absorption in barium-doped $BiFeO_3$ ceramics, J. Mater. Sci. **52**, 581 (2017).

[19] X. Huang, H. Yan, R. Harder, Y. Hwu, I. K. Robinson, and Y. S. Chu, Optimization of overlap uniformness for ptychography, Opt. Express **22**, 12634 (2014).

[20] K. Wakonig, H.-C. Stadler, M. Odstrčil, E. H. R. Tsai, A. Diaz, M. Holler, I. Usov, J. Raabe, A. Menzel, and M. Guizar-Sicairos, *PtychoShelves*, a versatile high-level framework for high-performance analysis of ptychographic data, J. Appl. Cryst. **53**, 574 (2020).

[21] P. Thibault, M. Dierolf, A. Menzel, O. Bunk, C. David, and F. Pfeiffer, High-resolution scanning X-ray diffraction microscopy, Science **321**, 379 (2008).

[22] P. Thibault and A. Menzel, Reconstructing state mixtures from diffraction measurements, Nature **494**, 68 (2013).

[23] Q. Shi, E. Parsonnet, X. Cheng, N. Fedorova, R.-C. Peng, A. Fernandez, A. Qualls, X. Huang, X. Chang, H. Zhang, D. Pesquera, S. Das, D. Nikonov, I. Young, L.-Q. Chen, L. W. Martin, Y.-L. Huang, J. Íñiguez, and R. Ramesh, The role of lattice dynamics in ferroelectric switching, Nat. Commun. **13**, 1110 (2022).

[24] D. D. Fong, A. M. Kolpak, J. A. Eastman, S. K. Streiffer, P. H. Fuoss, G. B. Stephenson, C. Thompson, D. M. Kim, K. J. Choi, C. B. Eom, I. Grinberg, and A. M. Rappe, Stabilization of monodomain polarization in ultrathin $PbTiO_3$ films, Phys. Rev. Lett. **96**, 127601 (2006).

[25] P. Meisenheimer, G. Moore, S. Zhou, H. Zhang, X. Huang, S. Husain, X. Chen, L. W. Martin, K. A. Persson, S. Griffin, L. Caretta, P. Stevenson, and R. Ramesh, Switching the spin cycloid in $BiFeO_3$ with an electric field, Nat. Commun. **15**, 2903 (2024).

[26] A. Haykal, J. Fischer, W. Akhtar, J.-Y. Chauleau, D. Sando, A. Finco, F. Godel, Y. A. Birkhölzer, C. Carrétéro, N. Jaouen, M. Bibes, M. Viret, S. Fusil, V. Jacques, and V. Garcia, Antiferromagnetic textures in $BiFeO_3$ controlled by strain and electric field, Nat. Commun. **11**, 1704 (2020).

[27] S. Husain *et al.*, Non-volatile magnon transport in a single domain multiferroic, Nat. Commun. **15**, 5966 (2024).